\documentclass[twocolumn,showpacs,preprintnumbers,amsmath,amssymb,prl]{revtex4-1}%
\usepackage{graphicx}% Include figure files
\usepackage{dcolumn}% Align table columns on decimal point
\usepackage{bm}% bold math
\usepackage{tabularx}% Table width
\usepackage{amsmath}% Higher math
\usepackage{textcomp}
\usepackage{upgreek}

\begin{document}

\title{Drastic pressure effect on the extremely large magnetoresistance in WTe$_2$: quantum oscillation study}
\author{P. L. Cai,$^1$ J. Hu,$^2$ L. P. He,$^1$ J. Pan,$^1$ X. C. Hong,$^1$ Z. Zhang,$^1$ J. Zhang,$^1$ J. Wei,$^2$ Z. Q. Mao,$^2$ and S. Y. Li$^{1,3,}$}
\email{shiyan$_$li@fudan.edu.cn}
\affiliation{$^1$State Key Laboratory of Surface Physics, Department of Physics, and Laboratory of Advanced Materials, Fudan University, Shanghai 200433, P. R. China\\
$^2$Department of Physics and Engineering Physics, Tulane University, New Orleans, Louisiana 70118, USA\\
$^3$Collaborative Innovation Center of Advanced Microstructures, Fudan University, Shanghai 200433, P. R. China}

\date{\today}

\begin{abstract}
The quantum oscillations of the magnetoresistance under ambient and high pressure have been studied for WTe$_2$ single crystals, in which extremely large magnetoresistance was discovered recently. By analyzing the Shubnikov-de Haas oscillations, four Fermi surfaces are identified, and two of them are found to persist to high pressure. The sizes of these two pockets are comparable, but show increasing difference with pressure. At 0.3 K and in 14.5 T, the magnetoresistance decreases drastically from 1.25 $\times$ $10^5$\% under ambient pressure to 7.47 $\times$ $10^3$\% under 23.6 kbar, which is likely caused by the relative change of Fermi surfaces. These results support the scenario that the perfect balance between the electron and hole populations is the origin of the extremely large magnetoresistance in WTe$_2$.

\end{abstract}


\maketitle

The magnetoresistance (MR) is an important transport property of condensed matters. For simple metals, the MR is usually very small, showing quadratic field dependence in low field and saturating in high field \cite{Pippard}. In contrast, giant magnetoresistance (GMR) was discovered in magnetic multilayers \cite{Baibichet}, and colossal magnetoresistance (CMR) was found in manganites \cite{Salamon}. Interestingly, in recent few years extremely large ($> 10^5$\%) and non-saturating positive MR (XMR) was discovered in some nonmagnetic compounds including PtSn$_4$ \cite{EMun}, Cd$_3$As$_2$ \cite{TLiang}, WTe$_2$ \cite{Ali}, and NbSb$_2$ \cite{KFWang}.

The very recent discovery of XMR in WTe$_2$ is of particular interest \cite{Ali}. WTe$_2$ is a layered transition-metal dichalcogenide, with the crystal structure shown in Fig. 1(a). In the dichalcogenide layers, W chains are formed along the $a$ axis. An XMR of 4.5 $\times$ $10^5$\% in 14.7 T at 4.5 K was found when the current is along the $a$ axis and magnetic field is applied along the $c$ axis \cite{Ali}. More remarkably, it increases to as high as 1.3 $\times$ $10^7$\% in 60 T at 0.53 K, without any sign of saturation \cite{Ali}. Such an XMR makes WTe$_2$ outstanding among transition-metal dichalcogenides, in which various interesting physics such as charge density wave and superconductivity have been extensively studied \cite{Morris,Moncton}.

Based on electronic structure calculations, the XMR in WTe$_2$ was attributed to the compensation between the electron and hole populations \cite{Ali}. Actually, such a two-band model for charge transport in semimetals was previously used to explain the very large MR observed in high-purity graphite and bismuth \cite{XDu}. However, in both graphite and Bi, the MR saturates beyond a few Tesla, due to the slight deviation from perfect compensation \cite{Kopelevich,Fauque,Ali}. In this sense, the non-saturating XMR in WTe$_2$ may provide the first example of perfectly balanced electron-hole populations \cite{Ali}. While the subsequent angle-resolved photoemission spectroscopy (ARPES) experiments observed one pair of hole and electron pockets of approximately the same size at low temperature \cite{TValla}, a very recent ARPES work showed a complex Fermi surface topology with two pairs of hole and electron pockets \cite{JJiang}, therefore more experiments like quantum oscillation measurement are highly desired to clarify this important issue. On the other hand, since the perfect balance between electron and hole populations should be very sensitive to some tuning parameters such as doping and pressure, investigating the evolution of the XMR in WTe$_2$ with these parameters may further clarify its origin.

In this Letter, we present the quantum oscillation study of the magnetoresistance for WTe$_2$ single crystals under ambient and high pressure. By analyzing the Shubnikov-de Haas oscillations of magnetoresistance at low temperature, four Fermi surfaces are revealed, which should correspond to the two pairs of hole and electron pockets later probed by ARPES experiments \cite{JJiang}. A drastic suppression of the XMR with increasing pressure is observed, which is accompanied by a change of the Fermi surface topology. The correlation between them provides support for the mechanism that the XMR of WTe$_2$ originates from the electron-hole compensation.

\begin{figure}
\centering
\centering
\includegraphics[clip,width=0.50\textwidth]{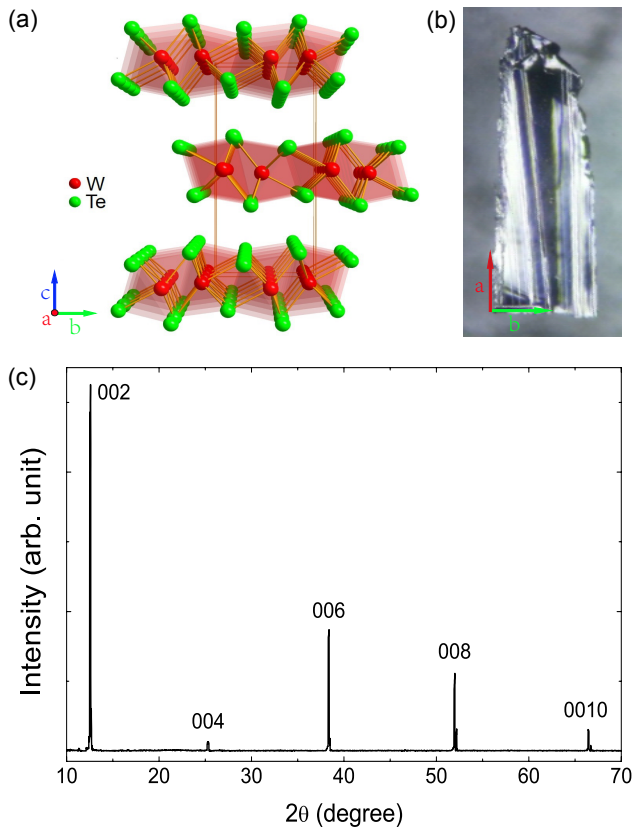}
\caption{(Color online) (a) Crystal structure of WTe$_2$: a three-dimensional perspective view along the $a$-axis direction. W chains are formed along the $a$ axis. (b) The photo of one cleaved WTe$_2$ sample. The stripes along the $a$ axis can be clearly seen. (c) The X-ray diffraction patten of WTe$_2$ single crystal. Only reflections of (0 0 2$l$) show up, indicating the $c$-axis orientation.}
\end{figure}

\begin{figure}
\centering
\centering
\includegraphics[clip,width=0.45\textwidth]{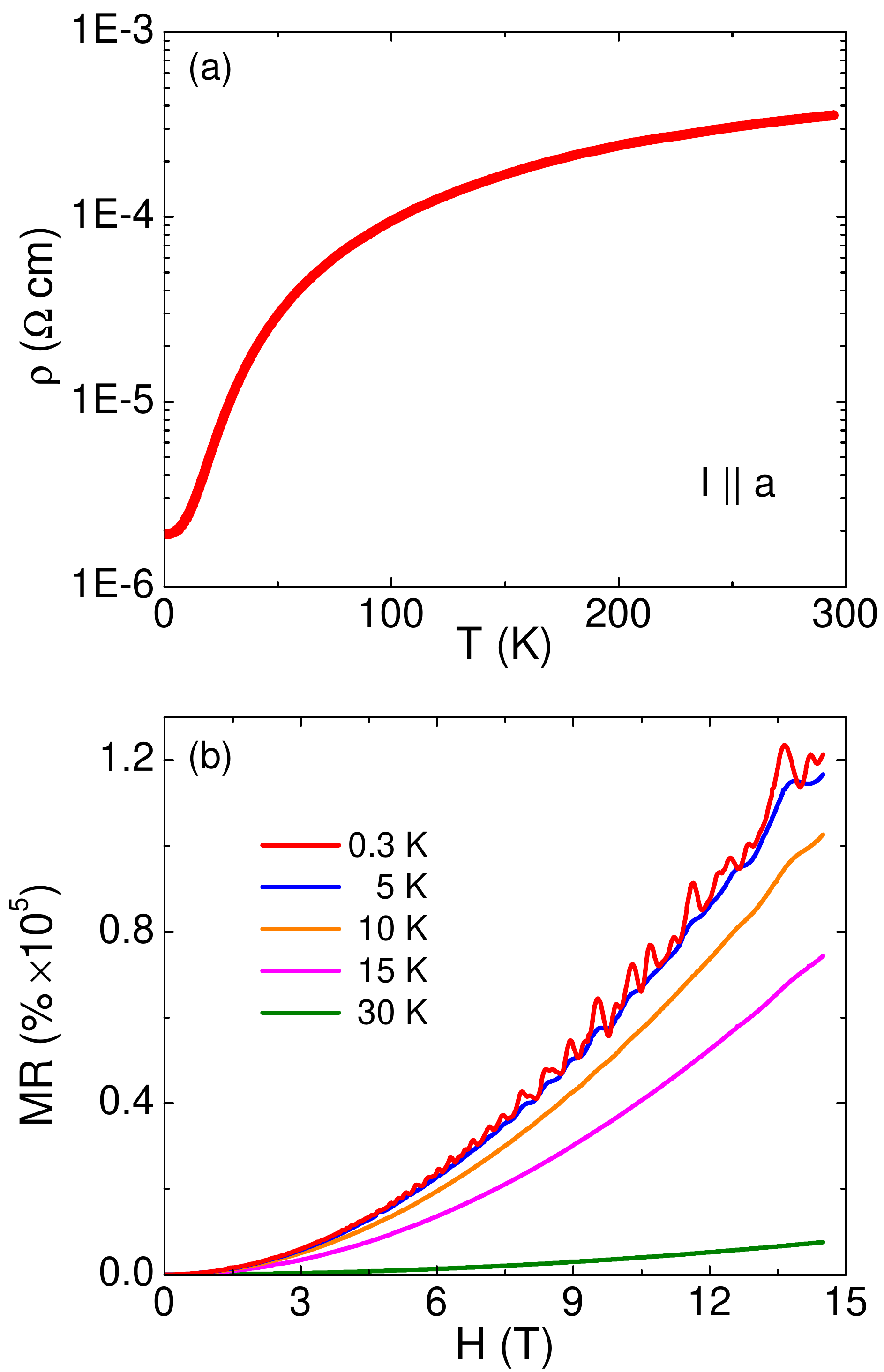}
\caption{(Color online) (a) Temperature dependence of resistivity in zero field, with current along the $a$ axis. (b) The magnetoresistance of WTe$_2$ single crystal up to 14.5 T at various temperatures, with magnetic field applied along the $c$-axis direction. There are clear Shubnikov-de Haas oscillations below 5 K.}
\end{figure}

The WTe$_2$ single crystals were grown by using a chemical vapor transport method similar to that described in Ref. \cite{Kabashima,Ali}. A mixture of stoichiometric W and Te powder were sealed into an evacuated quartz tube with transport agent Br$_2$. The quartz tube was then placed in a double zone furnace with temperature gradient of 100 $^\circ$C between 750 $^\circ$C and 650 $^\circ$C. Large single crystals of centimeter size were obtained after one week. The samples were cut and cleaved to a typical size of 2.0 $\times$ 0.7 $\times$ 0.02 mm$^3$, in which the longest edge is along $a$ axis and the thinnest dimension is along the $c$ axis. Figure 1(b) shows the photo of one sample. X-ray diffraction (XRD) measurement was performed using an X-ray diffractometer (D8 Advance, Bruker). Standard four-probe method was used for resistivity measurements, with current along the $a$ axis. The contacts were made with silver epoxy. The resistivity was measured in a $^4$He cryostat from 300 K to 2 K, and in a $^3$He cryostat down to 0.3 K. For measurements under pressure, samples were pressurized in a piston-cylinder clamp cell made of Be-Cu alloy, with Daphne oil as the pressure medium. The pressure inside the cell was determined from the $T_c$ of a tin wire. Magnetic field was applied along the $c$-axis direction up to 14.5 T.

Figure 1(c) presents the XRD result of WTe$_2$ single crystal. Only reflections of (0 0 2$l$) show up, indicating the $c$-axis orientation. The lattice parameter $c =$ 14.054 \AA\ is determined from the XRD data, which agrees well with previous reports \cite{Brown,AMar}.

Figure 2(a) shows the temperature dependence of resistivity in zero field for a WTe$_2$ single crystal, with current along the $a$ axis. It has $\rho$(295 K) = 355 $\mu$$\Omega$ cm and $\rho$(2 K) = 1.93 $\mu$$\Omega$ cm, with the residual resistivity ratio RRR = $\rho$(295 K)/$\rho$(2 K) = 184. This RRR value is about half of that in Ref. \cite{Ali}. Figure 2(b) presents the magnetoresistance up to 14.5 T at various temperatures. The MR is defined by MR = [$\rho$($H$) - $\rho$(0 T)]/$\rho$(0 T) $\times$ 100$\%$. At 0.3 K and in 14.5 T, the MR reaches as high as 1.25 $\times$ $10^5$\% (taken from the smooth background of the curve). This value is lower than that in Ref. \cite{Ali}, which is attributed to the slightly lower quality (smaller RRR) of our sample. With increasing temperature, the MR decreases rapidly and the oscillations disappear above 15 K, which are consistent with Ref. \cite{Ali}.

The SdH oscillation is a useful technique to detect the Fermi surface topology \cite{Shoenberg}. In Fig. 3(a), the oscillatory MR is analyzed by employing Fast Fourier Transform (FFT) for various temperatures from 0.3 to 6 K. The FFT spectrum shows four major peaks at 94.7, 132, 148, and 166 T oscillation frequency, labeled as $\gamma, \alpha, \beta,$ and $\delta$. The second harmonics 2$\alpha$ and 2$\beta$, likely due to spin-splitting, and the sum of $\alpha$ and $\beta$ due to magnetic breakdown are also observed \cite{Shoenberg}. The SdH oscillation frequency $F$ is proportional to the extremal cross-sectional area of the Fermi surface normal to the field, according to the Onsager relation $F$ = ($\Phi$$_0$/2$\pi$$^2$)$A_F$. The $\gamma, \alpha, \beta,$ and $\delta$ peaks seen in Fig. 3(a) clearly indicate that there exist four Fermi pockets normal to the field.

\begin{figure}
\centering
\includegraphics[clip,width=0.45\textwidth]{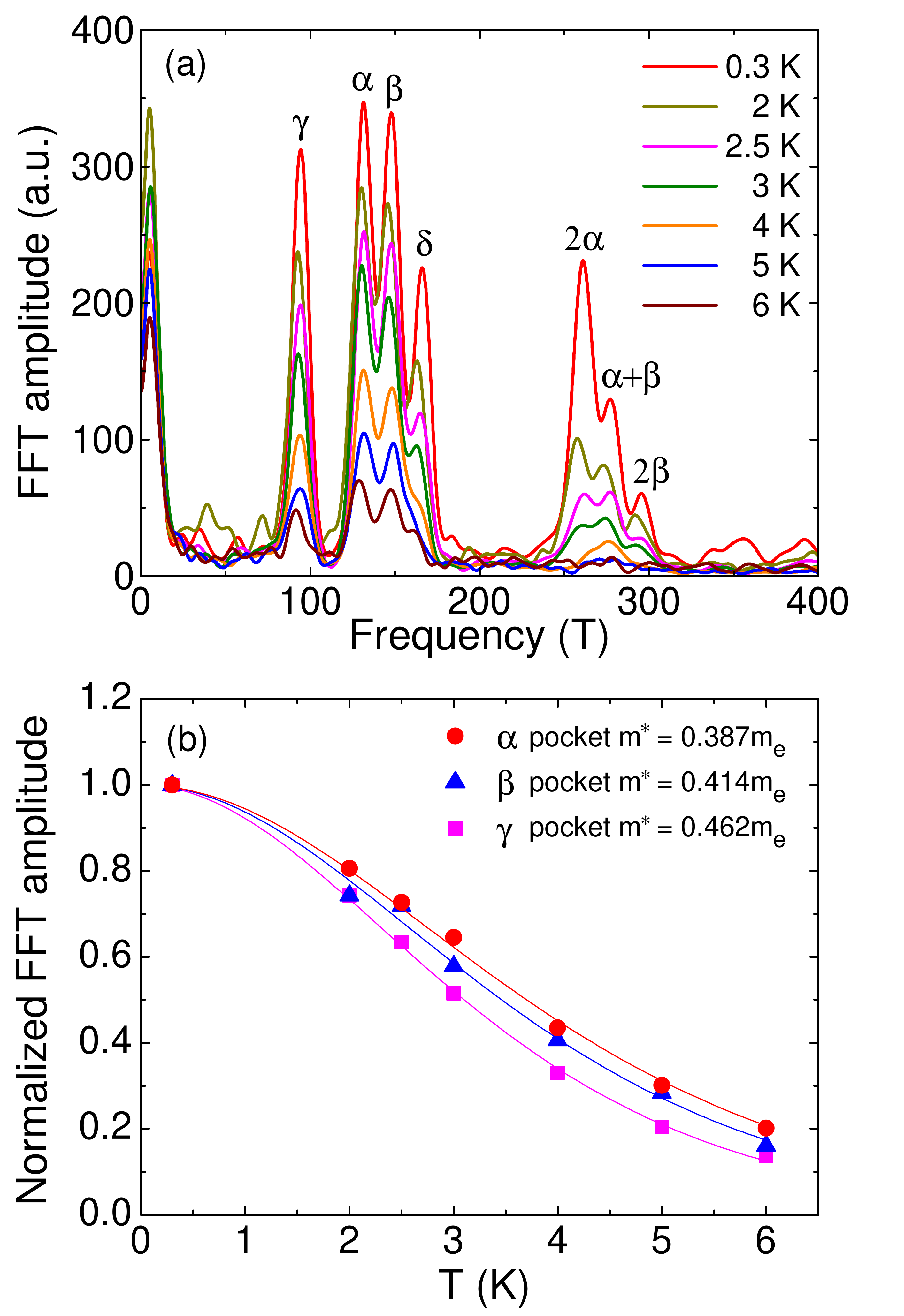}
\caption{(Color online) (a) The Fast Fourier Transform (FFT) of the oscillatory magnetoresistance shows four major peaks, labeled as $\alpha, \beta, \gamma$, and $\delta$. The second harmonics 2$\alpha$ and 2$\beta$, and the sum of $\alpha$ and $\beta$ are also observed. (b) The temperature dependence of FFT amplitude of $\alpha, \beta$, and $\gamma$ peaks, normalized by their 0.3 K values. The solid lines are the fit using the Lifshitz-Kosevich formula, which give the effective mass $m^*$ = 0.387$m_e$, 0.414$m_e$, and 0.462$m_e$, for $\alpha, \beta$, and $\gamma$ pockets, respectively.}
\end{figure}

\begin{figure}
\centering
\includegraphics[clip,width=0.45\textwidth]{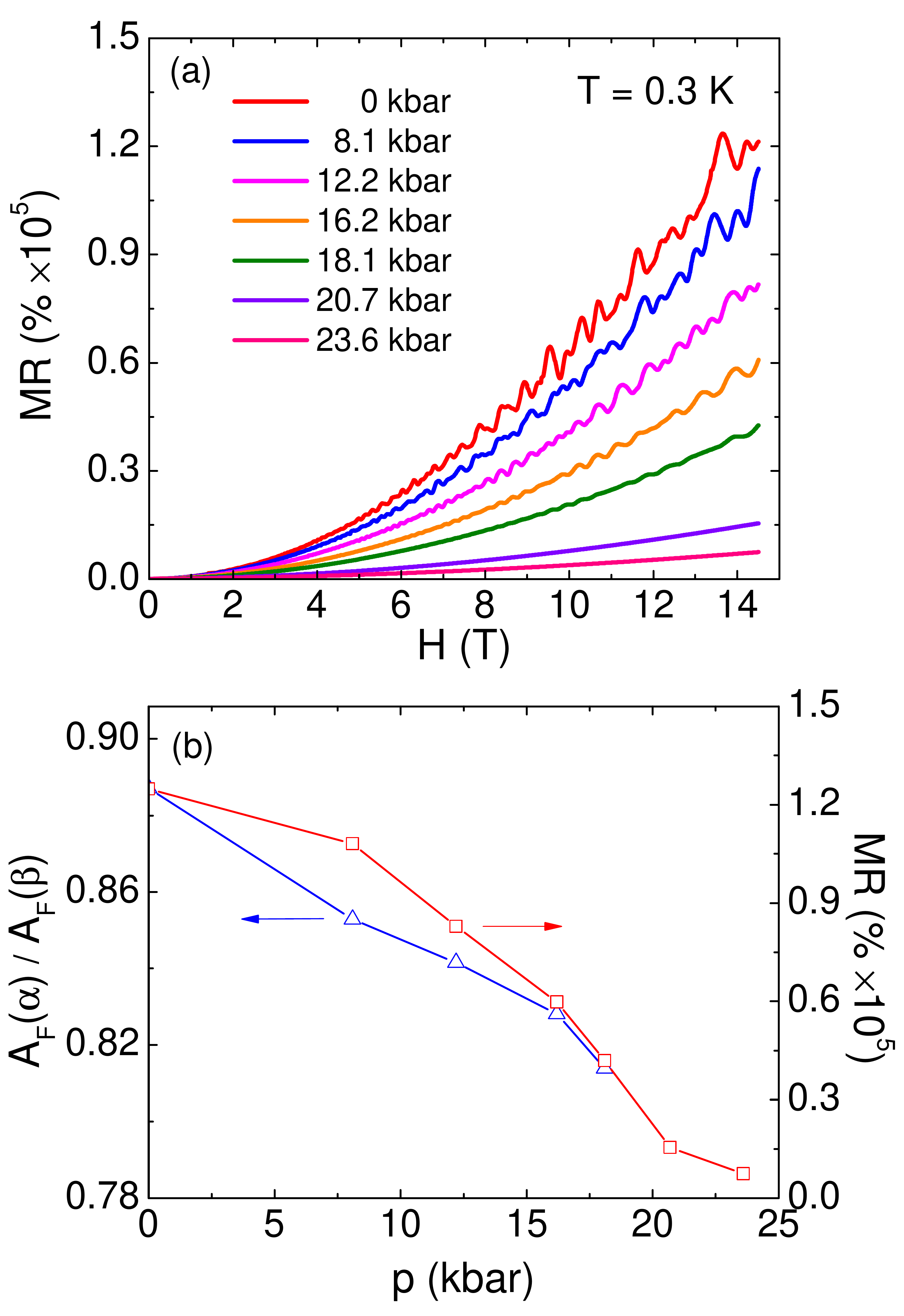}
\caption{(Color online) (a) The magnetoresistance of WTe$_2$ single crystal under various pressures up to 23.6 kbar, measured at $T$ = 0.3 K. (b) The pressure dependence of the magnetoresistance at $T$ = 0.3 K and in $H$ = 14.5 T. The data points are taken from the smooth background of each curve in (a). The magnetoresistance decreases from 1.25 $\times$ $10^5$\% under ambient pressure to 7.47 $\times$ $10^3$\% under $p$ = 23.6 kbar. For the pair of electron and hole pockets $\alpha$ and $\beta$, the ratio of their area $A_F(\alpha)/A_F(\beta)$ manifests similar pressure evolution (see text).}
\end{figure}

The SdH oscillations of MR can be described by the Lifshitz-Kosevich (LK) formula \cite{Shoenberg}. The temperature dependence of the oscillation amplitude is determined by the thermal damping factor $R_T$ in the LK formula, defined as $R_{T}$ = $\frac{\alpha Tm^{*}/B}{sinh(\alpha Tm^{*}/B)}$, where $T$ is the temperature, $m^{*}$ is the effective mass, 1/$B$ is the average inverse field of the Fourier window, $\alpha$ = $2\pi^{2}k_{B}/e\hbar$, and $k_B$ is the Boltzmann constant. Figure 3(b) shows the fits of the temperature dependence of the normalized oscillation amplitudes to the thermal damping factor $R_T$ from 0.3 to 6 K. The effective masses $m^{*}$ = 0.387$m_e$, 0.414$m_e$, and 0.462$m_e$ are obtained for $\alpha, \beta$, and $\gamma$ pockets, respectively, where $m_e$ is the bare electron mass. Since the amplitude of the $\delta$ peak can not be reliably extracted above 3 K, we did not do the fitting for the $\delta$ pocket.

Next we investigate the pressure effect on the XMR of WTe$_2$. Figure 4(a) shows the MR of WTe$_2$ single crystal under various pressures up to 23.6 kbar, measured at $T$ = 0.3 K. With increasing pressure, the MR is strongly suppressed and the oscillations also gradually disappear. The pressure dependence of the MR in the highest field $H$ = 14.5 T is plotted in Fig. 4(b). It decreases from 1.25 $\times$ $10^5$\% under ambient pressure to 7.47 $\times$ $10^3$\% under $p$ = 23.6 kbar.

To find out the cause of this strong suppression of MR in WTe$_2$, we examine the evolution of its Fermi surfaces with pressure. First let us identify the four Fermi pockets obtained at ambient pressure, by comparing with the two ARPES results \cite{TValla,JJiang}. One ARPES group found a pair of hole and electron pockets with comparable size along the $\Gamma-X$ direction \cite{TValla}, however, another ARPES group revealed a more complex Fermi surface topology with two pairs of hole and electron pockets \cite{JJiang}, as sketched in Fig. 5(a). The sizes of the pair of hole and electron pockets in Ref. \cite{TValla} are comparable with $k_F$ $\approx$ 0.08 \AA$^{-1}$, while the sizes of the two pairs of hole and electron pockets in Ref. \cite{JJiang} are smaller and show slight difference. We examine the sizes of the four Fermi pockets obtained from our SdH oscillations of MR. According to the Onsager relation $F$ = ($\Phi$$_0$/2$\pi$$^2$)$A_F$, $A_F$ = 0.0090, 0.0125, 0.0141, and 0.0158 \AA$^{-2}$ are obtained for $\gamma$, $\alpha$, $\beta$, and $\delta$ pockets, respectively. By assuming a circular pocket, the Fermi momentums $k_F(\gamma)$ $\approx$ 0.054 \AA$^{-1}$,  $k_F(\alpha)$ $\approx$ 0.063 \AA$^{-1}$, $k_F(\beta)$ $\approx$ 0.067 \AA$^{-1}$, and $k_F(\delta)$ $\approx$ 0.071 \AA$^{-1}$ are estimated. Therefore, both the number and sizes of the Fermi pockets obtained from our measurements are consistent with the Fermi surface topology of WTe$_2$ revealed in Ref. \cite{JJiang}. Note that an additional hole pocket around $\Gamma$ was observed in some samples but absent in some other samples \cite{JJiang}, therefore we do not sketch it in Fig. 5(a). It is likely absent in our samples.

\begin{figure}
\centering
\includegraphics[clip,width=0.43\textwidth]{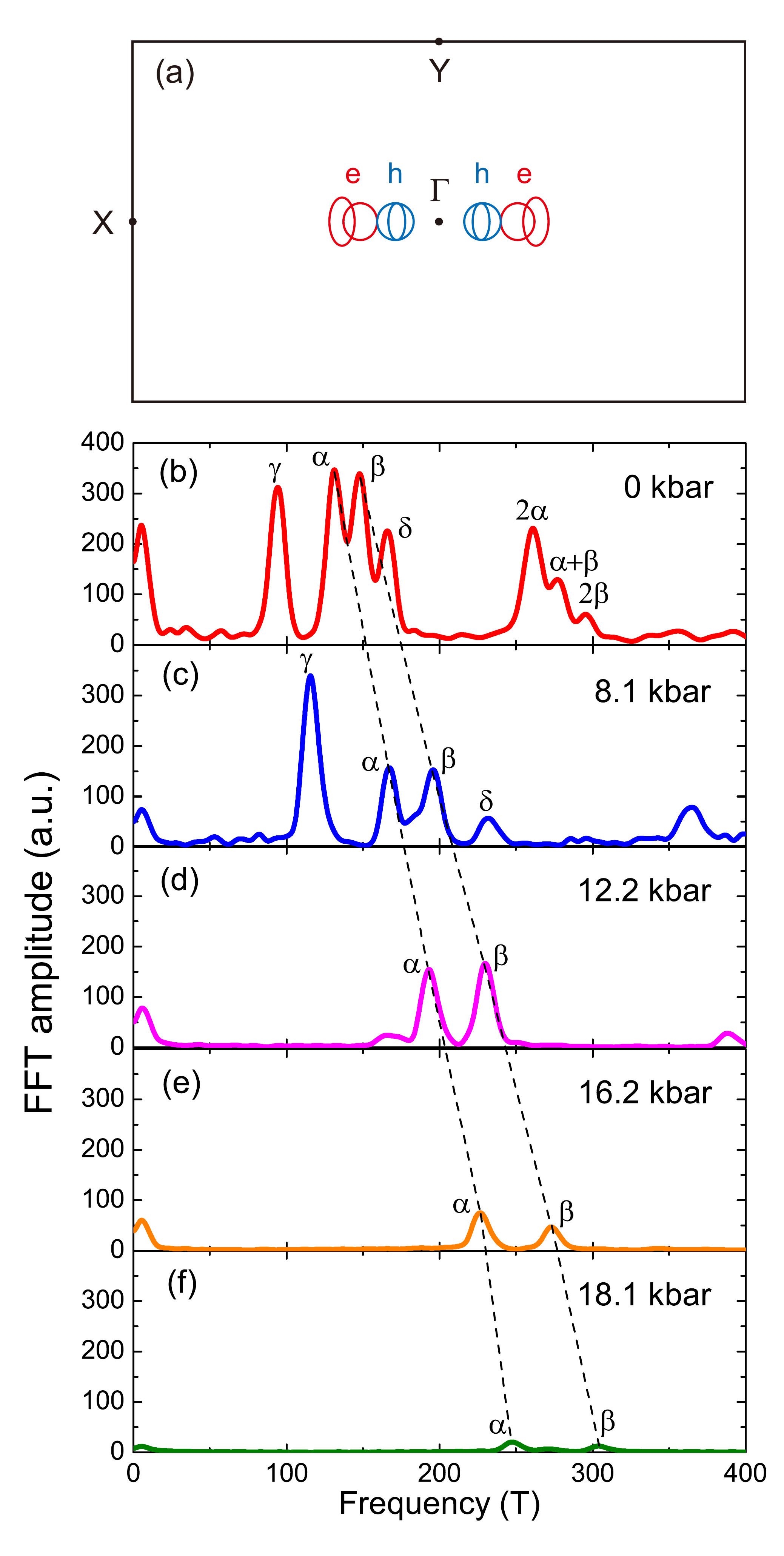}
\caption{(Color online) (a) Sketch of two pairs of hole and electron pockets along the $\Gamma-X$ direction in the Brillouin zone of WTe$_2$, as observed by ARPES experiments \cite{JJiang}. (b)-(f) The FFT spectrum of the oscillatory magnetoresistance under various pressures up to 18.1 kbar. The $\alpha$ and $\beta$ peaks persist to high pressure, which are identified as the pair of electron and hole pockets in (a). From (b) to (f), the dash lines are guide to the eye to show that the sizes of electron and hole pockets become increasingly different.}
\end{figure}

Figure 5(b)-(f) show the FFT spectrums from ambient pressure to 18.1 kbar. Above 18.1 kbar, the oscillations are too weak to give a reliable FFT spectrum. One can see that both $\alpha$ and $\beta$ peaks persist all the way to the highest pressure. According to the scenario in Ref. \cite{Ali}, the nearly perfect balance of electron and hole populations is responsible for the XMR in WTe$_2$. Since under 12.2 kbar, the MR is still as large as 8.28 $\times$ $10^4$\% and only the $\alpha$ and $\beta$ peaks remain, we identify $\alpha$ and $\beta$ as one of the two pairs of hole and electron pockets. It will be very interesting to do electronic structure calculation under pressure, to show how another pair of hole and electron pockets, $\gamma$ and $\delta$, disappears with increasing pressure. We note a recent theoretical calculation shows that the electronic structure of monolayer 1T'-WTe$_2$ is sensitive to the tensile strain, which may be crucial for realizing the quantum spin Hall effect in this two-dimensional transition metal dichalcogenide \cite{XFQian}.

With increasing pressure, the absolute sizes of $\alpha$ and $\beta$ pockets increase, because of the shrink of lattice. The relative size of $\alpha$ and $\beta$ pockets, defined as $A_F(\alpha)$/$A_F(\beta)$, is plotted in Fig. 4(b) together with the pressure dependence of MR. The two curves are clearly correlated, suggesting that the increasing difference between the sizes of $\alpha$ and $\beta$ pockets is the cause of the strong suppression of MR with pressure in WTe$_2$. In this sense, our result confirms the importance of the perfect balance between the electron and hole populations to the XMR in WTe$_2$.

Note that the four major frequencies we observed ($\gamma$ = 94.7 T, $\alpha$ = 132 T, $\beta$ = 148 T, and $\delta$ = 166 T) were confirmed by a later quantum oscillation study of WTe$_2$ \cite{ZWZhu}. According to the FFT analysis of the quantum oscillations of the Seebeck coefficient with field applied along the $c$ axis, they found four major frequencies $F^1$ = 92 T, $F^2$ = 125 T, $F^3$ = 142 T, and $F^4$ = 162 T \cite{ZWZhu}, which are nearly the same as ours. However, they identified $F^2$ and $F^3$ ($\alpha$ and $\beta$ in our Fig. 5) as two electron pockets (``Russian doll" structure), based on their band calculations \cite{ZWZhu}. In fact, the band calculations of WTe$_2$ are very subtle, in the number and size of electron and hole pockets \cite{Ali,ZWZhu,JJiang,HYLv,YFZhao}. The calculated electronic structure only partially reproduces the experimental bands and Fermi surface \cite{JJiang}, thus it is not appropriate to identify the four major frequencies only based on band calculations \cite{ZWZhu}. This situation highlights the importance of our pressure study, in which the large MR = 8.28 $\times$ $10^4$\% and the remaining $\alpha$ and $\beta$ peaks under 12.2 kbar enable us to identify $\alpha$ and $\beta$ ($F^2$ and $F^3$ in Ref. \cite{ZWZhu}) as one pair of electron and hole pockets. Therefore, the correct Fermi surface topology is very likely that revealed by the ARPES experiments in Ref. \cite{JJiang}, which is consistent with our results. Such a Fermi surface topology is the base to understand the XMR in WTe$_2$.

In summary, we study the quantum oscillations of magnetoresistance under ambient and high pressure for WTe$_2$ single crystals. Under ambient pressure, four Fermi surfaces are identified by analyzing the SdH oscillations, which are likely two pairs of hole and electron pockets along the $\Gamma-X$ direction. With increasing pressure, drastic change of Fermi surface topology and strong suppression of the XMR are observed. While one pair of hole and electron pockets ($\alpha$ and $\beta$) persists to high pressure, the other pair of hole and electron pockets ($\gamma$ and $\delta$) disappears with increasing pressure. The relative size of the $\alpha$ and $\beta$ pockets decreases with increasing pressure, which may cause the strong suppression of the XMR. Our results support the scenario that the perfect balance between the electron and hole populations is the origin of the XMR in WTe$_2$.

We thank J. K. Dong, D. L. Feng, J. Jiang, P. S. Wang, and Z. J. Xiang for helpful discussions. This work is supported by the Natural Science Foundation of China, the Ministry of Science and Technology of China (National Basic Research Program No. 2012CB821402 and 2015CB921401), Program for Professor of Special Appointment (Eastern Scholar) at Shanghai Institutions of Higher Learning, and STCSM of China (No. 15XD1500200). The work at Tulane is supported by NSF under Grant No. DMR-1205469.

{\it Note added:} After we put this work in arXiv (1412.8298), superconductivity was discovered in WTe$_2$ by applying higher pressure than ours \cite{FQSong,LLSun}, which coincides with the suppression of the XMR.

\end{document}